# Non Volatile MoS$_2$ Field Effect Transistors Directly Gated By Single Crystalline Epitaxial Ferroelectric

Zhongyuan Lu[1], Claudy Serrao[1], Asif Islam Khan[1], Long You[1], Justin C. Wong[1], Yu Ye[2], Hanyu Zhu[2], Xiang Zhang[2], and Sayeef Salahuddin[1, a)]

[1]Electrical Engineering and Computer Science, University of California, Berkeley, CA 94720, USA
[2]Mechanical Engineering, University of California, Berkeley, CA 94720, USA
a)email: sayeef@berkeley.edu

**Abstract:**

We demonstrate non-volatile, n-type, back-gated, MoS$_2$ transistors, placed directly on an epitaxial grown, single crystalline, PbZr$_{0.2}$Ti$_{0.8}$O$_3$ (PZT) ferroelectric. The transistors show decent ON current (19 µA/µm), high on-off ratio (10$^7$), and a subthreshold swing of (SS ~ 92 mV/dec) with a 100 nm thick PZT layer as the back gate oxide. Importantly, the ferroelectric polarization can directly control the channel charge, showing a clear anti-clockwise hysteresis. We have self-consistently confirmed the switching of the ferroelectric and corresponding change in channel current from a direct time-dependent measurement. Our results demonstrate that it is possible to obtain transistor operation directly on polar surfaces and therefore it should be possible to integrate 2D electronics with single crystalline functional oxides.

Single crystalline ferroelectric (FE) materials provide a wide range of functionalities including high speed of switching, large remnant polarization and therefore long memory retention time and also transduction to a wide range of stimuli such as pressure and temperature[1]. However, integrating the highest performance perovskite based FE materials with conventional electronics has traditionally been challenging due to significant lattice mismatch between Si and perovskites. In this context, two dimensional (2D) semiconductors[2–5] could be of great interest as it is possible to transfer and place them on arbitrary substrates and therefore it should be possible to directly integrate them with single crystalline ferroelectrics. Unfortunately, the efforts in this direction have been stymied significantly by the interface states that appear between the polar surface of ferroelectrics and the 2D layers. In fact, these surface states have such a large density that they completely screen out the polarization charge of the ferroelectric (typically $> 10^{14}/cm^2$), thereby decoupling the 2D layer from the ferroelectric. For example, a robust trap related hysteresis (often called the 'anti-hysteresis') has been observed in multiple studies[6–9]. A real ferroelectric induced hysteresis is rare[10–12]. Similarly, ferroelectric control of the channel charge for 2D transition metal dichalcogenides (TMDs) has proved to be significantly challenging[13–15]. While the 'anti-hysteresis' due to trapping/detrapping of interface traps have been shown to be very robust, this essentially eliminates all the functional properties offered by single crystal ferroelectrics. It has also been observed that 2D ferroelectric transistors showed no hysteresis, possibly because the ferroelectric film behaved mainly as a high-$\kappa$ dielectric layer[16]. Therefore, heterostructure of single crystal FE/2D channel materials where the transistor characteristic is controlled purely by the FE charge is desirable. In this paper, we demonstrate an N-type $MoS_2$ transistor with a single crystalline FE back-gate by directly transferring the $MoS_2$ layer on top of a single crystalline PZT thin film. The channel charge follows the FE polarization. We show that the interface trap states have a direct correlation with the surface quality of the FE material. Importantly, the transistors made of the transferred layers on PZT show excellent current-voltage characteristic, comparable to those obtained with standard high-$\kappa$ dielectric.

Approximately 100 nm thick single crystalline $PbZr_{0.2}Ti_{0.8}O_3$ (PZT) film was grown on an epitaxially matched $SrTiO_3$ (STO) (001) substrate via KrF pulsed laser deposition (PLD). A 30 nm $SrRuO_3$ (SRO) buffer layer was used between PZT and STO as the bottom electrode of the back-gate transistor structure. PZT and SRO were grown at 600°C and 700°C respectively with an oxygen background pressure of 100 mTorr. After film growth, the samples were cooled to room temperature in 1 atm oxygen at a rate of 10°C per minute. X-ray diffraction (XRD) analysis was used for phase identification (Figure 1a). Surface topography (Figure 1b) was measured using atomic force microscopy (AFM). The surface RMS roughness is ~0.478 nm. The polarization-voltage loops of the PZT capacitor are shown in Figure 1c. The permittivity vs voltage and admittance angle vs voltage behaviour are shown in Fig. 1d. The large remnant polarization, sharp switching and low leakage exemplified by high admittance angle indicate excellent electronic property of the synthesized film.

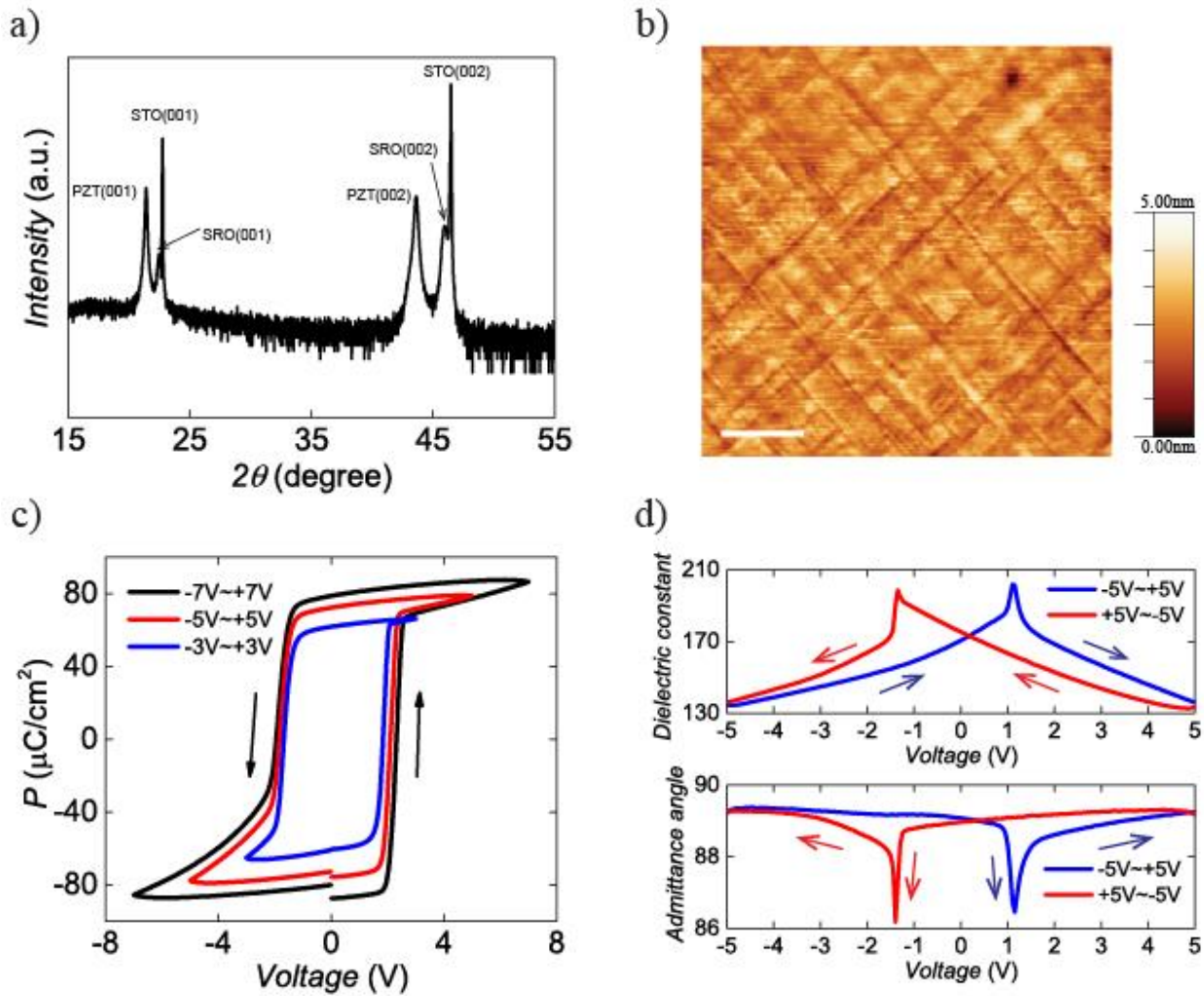

Figure 1. a) XRD pattern of PZT/SRO/STO structure. b) PZT film surface AFM topography. The scale bar is 1 µm. c) PZT film polarization-voltage loops for different voltage sweeping ranges. d) *C-V* characteristics of the PZT film.

Figure 2a shows a schematic of the back-gate MoS$_2$ transistor. MoS$_2$ flakes were mechanically exfoliated from bulk crystals onto 285 nm SiO$_2$/Si substrates (Figure 2b), which are optimal for estimating flake thickness via color contrast[17]. Multilayer MoS$_2$ flakes were chosen for high channel current. The selected MoS$_2$ flakes were transferred onto PZT substrates via dry cutting transfer process[18]. AFM was used to measure the thickness and uniformity of the flakes after transfer. No ripples or ruptures were found as shown in Figure 2c. The measured thickness of the transferred flake is ~6.81 nm (Figure 2d). Next e-beam lithography and metal evaporation were used to pattern 100 nm of Au film as the source/drain electrodes (Figure 2e). The channel is 3 µm in width and 5 µm in length. After lift-off, the device was annealed at 200°C in vacuum for 1 hour to remove adsorbates from the surface and reduce contact resistance[19,20]. An Agilent B1500A was used for current-voltage and current-time measurements, and all measurements were carried out in a high vacuum environment (2×10$^{-6}$ Torr). Note that we define the polarization direction as positive when it points into the channel and negative when it points out of the channel.

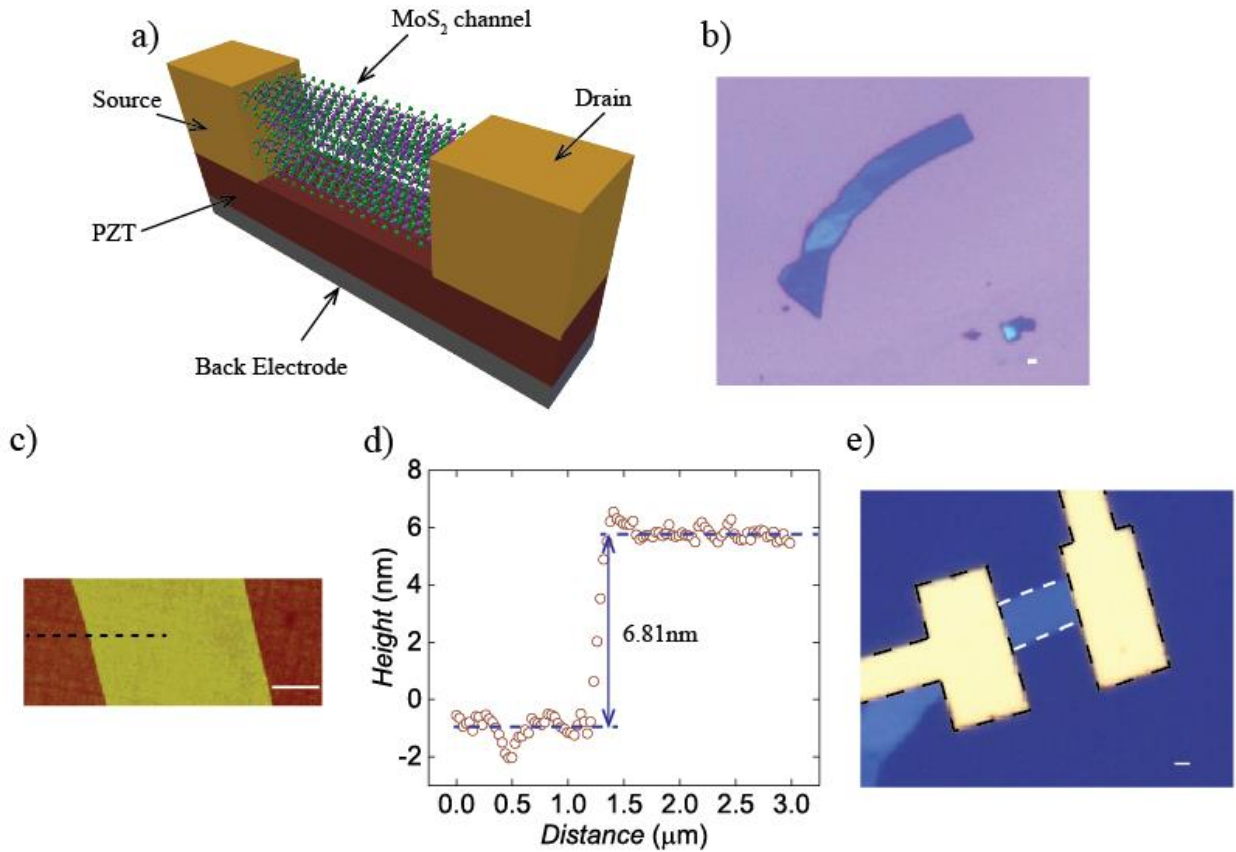

Figure 2. a) Schematic of multilayer MoS$_2$/PZT/SRO back-gate transistor. b) Optical image of MoS$_2$ flake exfoliated onto 285 nm SiO$_2$/Si substrate prior to transfer. c) AFM image of MoS$_2$ flake. d) AFM result of MoS$_2$ flake thickness. e) Optical image of the back-gate transistor. All scale bars are 1 μm.

The operating principle of our ferroelectric field-effect transistor (FeFET) is depicted in Figure 3a. In the traditional metal-oxide-semiconductor FET structure, applying a voltage across the gate oxide electrostatically dopes the channel via capacitive coupling. Conventional oxides can only exhibit polarization proportional to the applied voltage. In FeFETs, however, the ferroelectric maintains a remnant polarization that switches directions when an opposing electric field is applied with magnitude greater than the coercive electric field $E_c$. When the polarization is pointing towards the channel, it will induce electrons inn the MoS$_2$ layer, leading to a left-ward shift of the threshold voltage ($\Delta V_t < 0$); by contrast, when the polarization is pointing away from the channel, a rightward shift in the threshold voltage will ensue ($\Delta V_t > 0$). Thus for an n-type transistor, a counterclockwise hysteresis is expected. The two possible polarization states can represent the "1" and "0" states in non-volatile memory. In the transistor transfer curves (Figure 3b), the ON current reached as high as 19 μA/μm. Current saturation and a distinct anticlockwise hysteresis window were detected at positive gate voltage. Curves in different scales are overlapping each other, showing that surface states induced by adsorbates (e.g. O$_2$, H$_2$O, etc.) have minimal contribution at best. For the backward direction of the gate voltage sweep, the saturation current induced by positive polarization increases with wider sweep range. This indicates that the polarization amplitude increases with increasing voltage across the ferroelectric.

This trend is consistent with the polarization-electric field (PE) results in Figure 1c. As shown in Figure 3c, the ON/OFF ratio and subthreshold swing of the transistor reached ~$10^7$ and 92 mV/dec respectively, which are comparable to those of previously reported high-quality MoS$_2$ devices[3,4]. Unlike conventional FeFETs, there is no forward shift in threshold voltage $V_t$, implying that the negative polarization was ineffective. This is likely due to the lack of free electrons in the MoS$_2$ channel's depletion region, resulting in a weaker electric field across the ferroelectric film that is not strong enough to switch the polarization. This indicates that the observed effect of ferroelectric hysteresis is originating most likely from a partial switching of the ferroelectric. The red curves of Figure 3b shows changes in current direction as the gate voltage is swept from −3 V to +3 V. When the drain voltage $V_d$ is 1 V, the hysteresis loop is partially clockwise and partially anticlockwise. This occurs because the maximum voltage across the ferroelectric at the drain side can only reach $V_g − V_d = 2$ V while the maximum voltage at the source side is $V_g = 3$ V. Consequently, the electric field is not strong enough to switch the ferroelectric at the drain side. In contrast, when $V_d = 0.2$ V, the hysteresis loop is anticlockwise because the ferroelectric polarization state is the same at both the source and drain sides. Based on this analysis, we can conclude that the coercive voltage of the PZT film in the positive direction must be between 2.0 V and 2.8 V. This is reasonably consistent with PE measurements (Figure 1c).

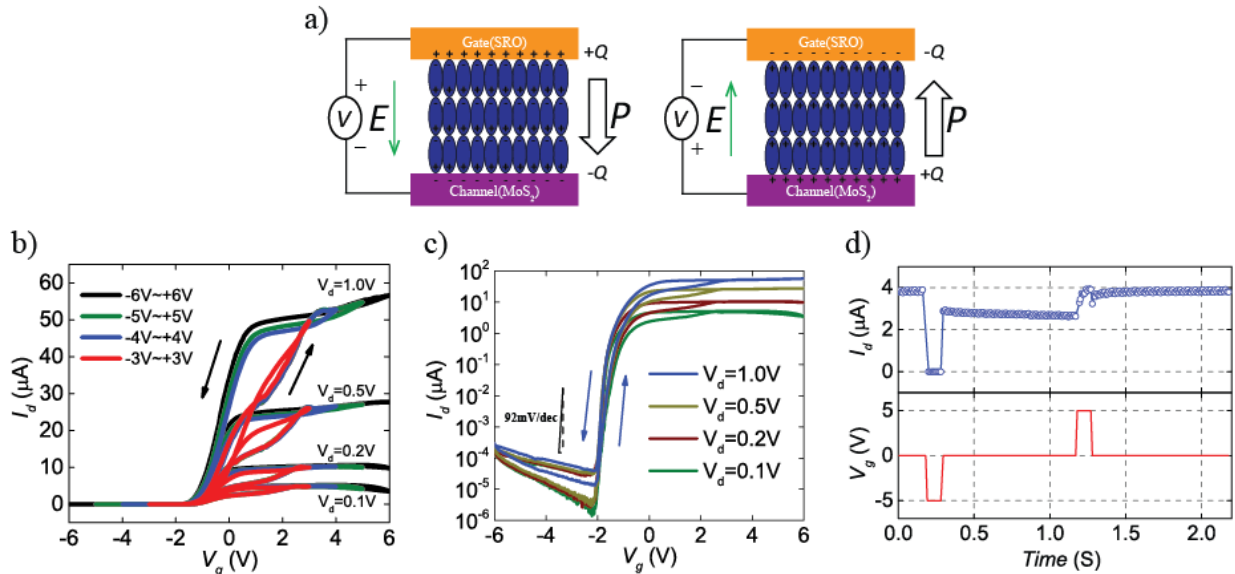

Figure 3. a) Schematic of polarization doping in the channel with different electric field directions. b) $I_d$-$V_g$ curves of MoS$_2$/PZT transistor with linear y-axis. c) $I_d$-$V_g$ curves of MoS$_2$/PZT transistor with logarithmic y-axis. d) Time-resolved channel current changes in response to gate voltage pulse sequences. $V_d = 0.1$V.

To test the polarization switching in-situ, we performed a time-resolved current measurement with a series of gate pulses. The red curve in Figure 3d shows the sequence of the applied gate pulses. The pulse has an amplitude of 5 V, pulse width of $t_{on} = 100$ ms, and pulse period of $t_{off} = 1$ s. The corresponding drain current is shown in the upper panel. First a negative pulse is applied. The current goes down to very small levels but comes back up when the voltage is turned OFF. However, the current does not come up to the same level as it was before applying the pulse. This indicates that a partial switching of ferroelectric is happening in the negative direction. Next a positive pulse is applied. The corresponding jump up in the current is clearly visible. The up directed change in current confirms the control of the channel charge by the polarization. The current is retained after the voltage is put back to zero. This shows that a robust polarization switching happens in our device in the positive direction. Together with the partial switching in the negative direction, this provides a complete memory operation for the device. Notably, the partial switching in the negative direction is expected because the channel is depleted for negative voltage and therefore most of the applied gate voltage drops across the channel capacitance rather than the ferroelectric. We have also performed a temperature dependent conductance measurement (see Fig. S2) and found that after the application of the positive voltage, the material is in the metallic phase as it would be expected from the polarity of the polarization.

In order to understand the effect of surface roughness on the transistor behavior, we have chosen an area of the PZT sample where the surface is relatively rough (see Fig. 4a). The polarization-voltage scan, as plotted in Fig. 4b, still shows excellent ferroelectric behavior. The $I_d$ - $V_g$ characteristic of a transistor fabricated on such a topography was measured in vacuum ($2\times10^{-6}$ Torr) avoid the effects from adsorbates. Figure 4c directly compares the transfer curves of the MoS$_2$ ferroelectric transistors fabricated on smooth and rough surfaces, showing that transistors fabricated on rough surfaces have hysteresis loops in the clockwise direction (i.e. anti-hysteresis), similar to those results reported before, not only of MoS$_2$ but also of graphene[6-9, 13-15]. Since the only difference between those two systems having different loop directions are the surface topography of PZT gate oxide layers, we postulate that a dominant reason for anti-hysteresis is interface states induced by defects on a rough surface.

Epitaxial ferroelectric films typically show excellent retention. Given that the transistors presented in this work are back gated, the retention of the individual transistors will be determined by the film itself. However, it is important to note that in a memory array, the retention will be ultimately determined by many factors other than the film itself, including stress cycles, operating temperature and READ disturb.

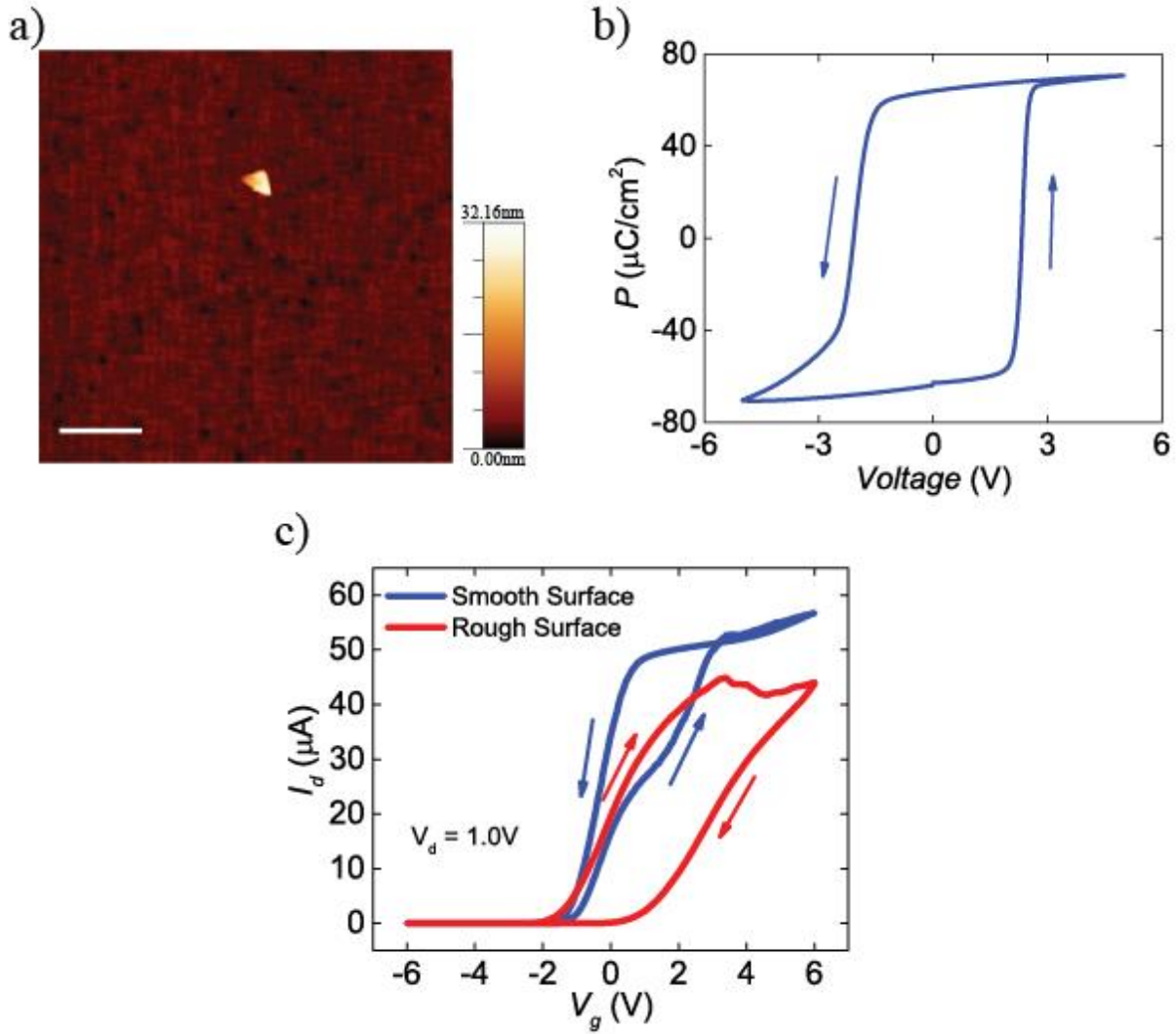

Figure 4. a) AFM topography of a rough PZT film. The scale bar is 1μm. b) The polarization-electric field loop of the rough PZT film. c) The transfer characteristics of MoS$_2$ transistors fabricated on PZT films with different surface qualities.

To summarize, we have fabricated n-type MoS$_2$ devices directly on single crystalline ferroelectric substrates. Our fabricated devices show excellent control of the channel charge from the ferroelectric polarization. A clear memory behavior is demonstrated. We show that a pristine and smooth surface is critical to ferroelectric control; otherwise the interface gets contaminated by surface charges which completely screens out the polarization, leading to clockwise hysteresis loop. Our work may be useful for non-volatile memory devices as well as integrating amplifiers directly on functional ferroelectric and piezoelectric oxides. In addition, it may be possible to use the large polarization of a single crystalline ferroelectric oxide to electronically induce metal insulator phase transition[20–23].

**References:**

[1] G.H. Haertling, J. Am. Ceram. Soc. **82**, 797 (1999).
[2] Q.H. Wang, K. Kalantar-Zadeh, A. Kis, J.N. Coleman, and M.S. Strano, Nat. Nanotechnol. **7**, 699 (2012).
[3] B. Radisavljevic, A. Radenovic, J. Brivio, V. Giacometti, and A. Kis, Nat. Nanotechnol. **6**, 147 (2011).
[4] S. Kim, A. Konar, W.-S. Hwang, J.H. Lee, J. Lee, J. Yang, C. Jung, H. Kim, J.-B. Yoo, J.-Y. Choi, Y.W. Jin, S.Y. Lee, D. Jena, W. Choi, and K. Kim, Nat. Commun. **3**, 1011 (2012).
[5] Y. Yoon, K. Ganapathi, and S. Salahuddin, Nano Lett. **11**, 3768 (2011).
[6] M. V. Strikha, JETP Lett. **95**, 198 (2012).
[7] E.B. Song, B. Lian, S. Min Kim, S. Lee, T.K. Chung, M. Wang, C. Zeng, G. Xu, K. Wong, Y. Zhou, H.I. Rasool, D.H. Seo, H.J. Chung, J. Heo, S. Seo, and K.L. Wang, Appl. Phys. Lett. **99**, 1 (2011).
[8] X. Hong, J. Hoffman, A. Posadas, K. Zou, C.H. Ahn, and J. Zhu, Appl. Phys. Lett. **97**, (2010).
[9] N. Park, H. Kang, J. Park, Y. Lee, Y. Yun, J. Lee, S. Lee, Y.H. Lee, and D. Suh, ACS Nano 10729 (2015).
[10] H.S. Lee, S.W. Min, M.K. Park, Y.T. Lee, P.J. Jeon, J.H. Kim, S. Ryu, and S. Im, Small **8**, 3111 (2012).
[11] C. Ko, Y. Lee, Y. Chen, J. Suh, D. Fu, A. Suslu, S. Lee, J.D. Clarkson, H.S. Choe, S. Tongay, R. Ramesh, and J. Wu, Adv. Mater. **28**, 2923 (2016).
[12] F.A. McGuire, Z. Cheng, K. Price, and A.D. Franklin, Appl. Phys. Lett. **109**, 1 (2016).
[13] X.W. Zhang, D. Xie, J.L. Xu, Y.L. Sun, X. Li, C. Zhang, R.X. Dai, Y.F. Zhao, X.M. Li, X. Li, and H.W. Zhu, IEEE Electron Device Lett. **36**, 784 (2015).
[14] A. Lipatov, P. Sharma, A. Gruverman, and A. Sinitskii, ACS Nano **9**, 8089 (2015).
[15] A. Nguyen, P. Sharma, T. Scott, E. Preciado, V. Klee, D. Sun, I.H. Lu, D. Barroso, S. Kim, V.Y. Shur, A.R. Akhmatkhanov, A. Gruverman, L. Bartels, and P.A. Dowben, Nano Lett. **15**, 3364 (2015).
[16] C. Zhou, X. Wang, S. Raju, Z. Lin, D. Villaroman, B. Huang, H.L.-W. Chan, M. Chan, and Y. Chai, Nanoscale **7**, 8695 (2015).
[17] M.M. Benameur, B. Radisavljevic, J.S. Héron, S. Sahoo, H. Berger, and A. Kis, Nanotechnology **22**, 125706 (2011).
[18] Y. Ye, Z.J. Wong, X. Lu, X. Ni, H. Zhu, X. Chen, Y. Wang, and X. Zhang, Nat. Photonics **9**, 733 (2015).
[19] H. Qiu, L. Pan, Z. Yao, J. Li, Y. Shi, and X. Wang, Appl. Phys. Lett. **100**, (2012).
[20] B.W.H. Baugher, H.O.H. Churchill, Y. Yang, and P. Jarillo-herrero, Nano Lett. **13**, 4212−4216 (2013).
[21] X. Chen, Z. Wu, S. Xu, L. Wang, R. Huang, Y. Han, W. Ye, W. Xiong, T. Han, G. Long, Y. Wang, Y. He, Y. Cai, P. Sheng, and N. Wang, Nat. Commun. **6**, 6088 (2015).
[22] B. Radisavljevic and A. Kis, Nat. Mater. **12**, 815 (2013).
[23] J.T. Ye, Y.J. Zhang, R. Akashi, M.S. Bahramy, R. Arita, and Y. Iwasa, Science (80-. ). **338**, 1193 (2012).

# Non Volatile MoS$_2$ Field Effect Transistors Directly Gated By Single Crystalline Epitaxial Ferroelectric

Zhongyuan Lu[1], Claudy Serrao[1], Asif Islam Khan[1], Long You[1], Justin C. Wong[1], Yu Ye[2], Hanyu Zhu[2], Xiang Zhang[2], and Sayeef Salahuddin[1, a)]

[1]Electrical Engineering and Computer Science, University of California, Berkeley, CA 94720, USA
[2]Mechanical Engineering, University of California, Berkeley, CA 94720, USA
a)email: sayeef@berkeley.edu

Common 2H-MoS$_2$ behaves as an insulator with conductivity that has a positive temperature coefficient due to the thermal excitation effect on carrier density. When the carrier density is low, transport is dominated by hopping between localized states induced by sulfur vacancies in the MoS$_2$ crystal[S1]. With high carrier density, MoS$_2$ changes to a metallic state in which strong coulomb interactions contribute to the conductivity[21]. For the ferroelectric transistor in the text with anticlockwise loops, since the remnant polarization of the PZT film pointing into the channel is stable, a large amount of electron doping to the MoS$_2$ channel was expected after a large enough positive voltage pulse generated on the gate. So an Agilent 81150A function generator was used to apply a +8 V voltage pulse ($t_{on}$ = 1ms) to the gate to fully polarize the ferroelectric gate oxide at different temperature points, and then the source/drain output performance was measured with the gate floating. In Figure S1a, the linear output curves represent ohmic contacts between source/drain and the channel even at 77 K. This implies that contact resistance is negligible and the channel resistance is nearly equal to the source/drain resistance. So the channel conductance could be directly calculated by:

$$G_{MoS_2} = \frac{I_d}{V_d}$$

where $I_d$ and $V_d$ are the channel current and source/drain voltage bias respectively. In Figure S1b, the channel conductance increases with decreasing temperature, meaning that MoS$_2$ is in the metallic phase with electrostatic doping from the positive remnant polarization.

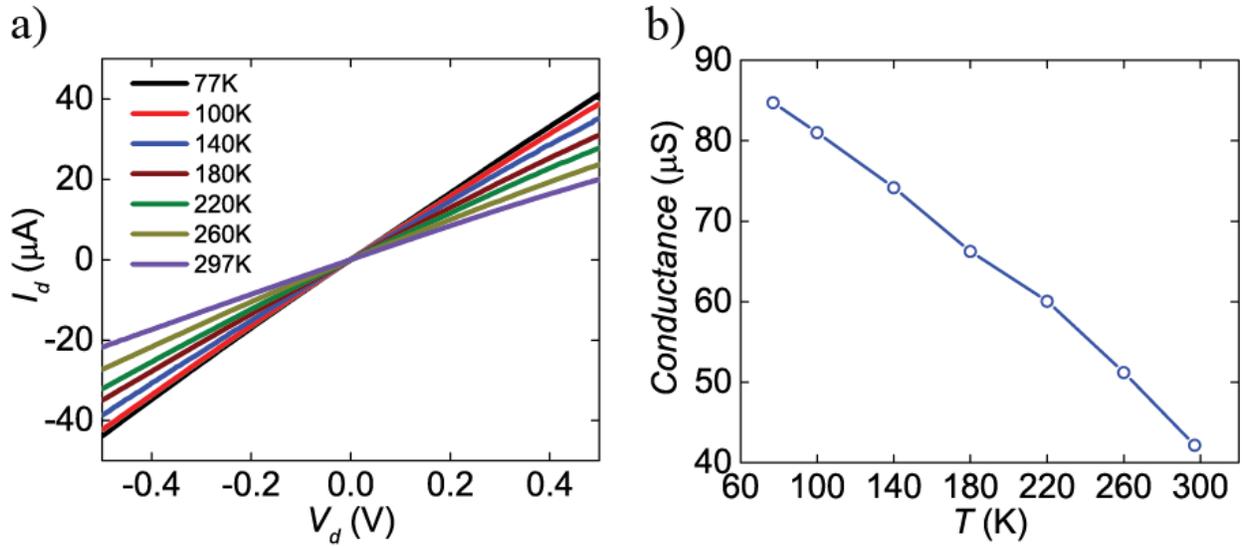

Figure S1. a) 2-terminal source/drain current at different temperatures after applying +8V positive gate voltage pulse (pulse width = 1ms). b) Channel conductance values at different temperatures.

**References:**

[S1] H. Qiu, T. Xu, Z. Wang, W. Ren, H. Nan, Z. Ni, Q. Chen, S. Yuan, F. Miao, F. Song, G. Long, Y. Shi, L. Sun, J. Wang, and X. Wang, Nat. Commun. **4**, 1 (2013).